\documentclass[structabstract]{aa}  
\usepackage{graphicx}
\usepackage{txfonts}
\usepackage{longtable,lscape}
\begin{document}
   \title{A spectroscopic survey of faint, high-Galactic-latitude red clump stars.
          I. The high resolution sample}

   \author{Marica Valentini
          \inst{1,2,3}
          \and
          Ulisse Munari \inst{1}
          }

   \institute{INAF-OAPD- Osservatorio Astronomico di Padova, via dell'Osservatorio 8,I-36012 Asiago (VI), Italy\\
              \email{marica.valentini@oapd.inaf.it, valentini@astro.ulg.ac.be}
         \and
             Dipartimento di Astronomia, Universit\'{a} di Padova, vicolo dell'Osservatorio 3,I-35122 Padova, Italy\\
              \email{ulisse.munari@oapd.inaf.it}
	 \and
             Institut d'Astrophysique et de G\'{e}ophysique de l'Universit\'{e} de Li\`{e}ge, All\'{e}e du 6 Ao\^{u}t 17, 4000 Li\`{e}ge, Belgium\\
             }

   \date{Received ......, 2010; accepted ....., 2010}

  \abstract
   {Their high intrinsic brightness and small dispersion in absolute
    magnitude make red clump (RC) stars a prime tracer of Galactic
    structure and kinematics.}
   {We aim to derive accurate, multi-epoch radial velocities and
    atmospheric parameters ($T_{\rm eff}$, $\log g$, [M/H], $V_{\rm rot}\sin
    i $) of a large sample of carefully selected RC stars, fainter than
    those present in other spectroscopic surveys and located over a great
    circle at high Galactic latitudes.}
   {We acquired data of the program stars of high signal-to-noise ratio (S/N) and 
    high resolution with
    the Asiago Echelle spectrograph.  Radial velocities were obtained 
    by applying cross-correlation and atmospheric parameters via $\chi^2$ fit to a
    synthetic spectral library.  Extensive tests were carried out by
    re-observing with the same instrument a large number of
    standard stars taken from a variety of sources in the literature.
    During these tests, we found that the absolute
    Tycho $V_{\rm T}$ magnitude of local red clump stars is not
    dependent on metallicity} 
   {A total of 277 red clump stars (101 of them with a second epoch
    observation) of the extended solar neighborhood and 55 calibration stars
    were observed and included in an output catalog that contains (in
    addition to relevant support astrometric and photometric data taken from
    literature) the main output of our survey: accurate multi-epoch radial
    velocities ($\sigma$(RV)$_\odot$$\leq$ 0.4 km/s), accurate atmospheric
    parameters ($\sigma$($T_{\rm eff}$)=68 K, $\sigma$($\log g$)=0.11 dex,
    $\sigma$([M/H])=0.10 dex, $\sigma$($V_{\rm rot}\sin i $)=1.1 km/sec),
    distances, and space velocities (U,V,W).}
   {}  

   \titlerunning{Spectroscopy of faint, high-Galactic-latitude red clump stars}
   \maketitle

\section{Introduction}

Since the pioneering work of Cannon (1970) and Faulkner and Cannon (1973),
the clump on the red giant branch of open clusters has been recognized as having been
formed by stars in the stage of central helium burning.  These authors interpreted
the near constancy of the clump absolute magnitude as the result of He
ignition in an electron-degenerate core.  Under these conditions, He burning
cannot begin until the stellar core mass attains a critical value of about
0.45 M$_{\odot}$.  It then follows that low-mass stars developing a
degenerate He core after H exhaustion, have similar core masses at the
beginning of He burning, and hence a similar luminosity.  A great contribution
to the study of red clump stars (hereafter RC stars) came from the
availability of Hipparcos parallaxes (Perryman et al.  1997).  On the CMD of
nearby stars built from precise Hipparcos parallaxes, the red clump is a
prominent and well populated feature.  The Hipparcos catalog (ESA 1997)
contained $\sim$600 RC stars with a parallax error $\sigma_\pi /
\pi$$\leq$10\% (Girardi et al. 1998), a wealth of data that made it possible to
transform the RC stars into calibrated standard candles.
 
Several calibrations of the absolute magnitude of RC stars have been carried
out.  Keenan \& Barnbaum (1999) found the absolute magnitude of the RC stars
in the optical to range from $M_V$=$+$0.70 for those of spectral type G8~III
to $M_V$=$+$1.00 for the K2~III ones.  From a sample of 228 RC stars
(satisfying the selection conditions $\sigma_\pi / \pi$$\leq$10\%,
+0.8$\leq$$V$$-$$I$$\leq$+1.25, and $-$1.4$\leq$$M_I$$\leq$+1.2 mag), Paczynsky
and Stanek (1998) found $ M_{I}=-0.28 \pm 0.09$, with no significant
dependence on metallicity.  Adopting the same selection criteria for a
larger sample of $\sim$600 RC stars observed by Hipparcos, Stanek and
Garnavich (1998) obtained $ M_{I}=-0.23 \pm 0.03$ and confirmed the
absence of any dependence on metallicity. Udalski (2000), using the same
selection conditions as Paczynsky and Stanek (1998), found instead a weak
dependence on metallicity in the form $M_{I}$=$-0.26 (\pm 0.02) + (0.13 \pm
0.07)([Fe/H]+0.25)$.  Alves (2000) found $M_{K}$=$-1.61 \pm 0.03$, with a
negligible dependence on metallicity in the form $ M_{K}$=$-1.64 (\pm 0.07)
+ (0.57 \pm 0.36)([Fe/H]+0.25)$.  Groenewegen(2008), working this time with
the revised Hipparcos parallaxes of van Leeuwen (2007), found slightly
fainter absolute magnitudes for RC stars, $M_{I}$=$-0.22 (\pm 0.03)$, and
$M_{K}$=$-1.54 (\pm 0.04)$, confirming a null or very weak dependence on
metallicity.  These calibrations imply that the absolute magnitude of the
red clump of the solar neighborhood has a rather small (or null) dependence
on metallicity.  However, Sarajedini (1999) and Twarog et al.  (1999),
working on open clusters, found some dependence of the absolute magnitude of
RC on metallicity and age, which is supported by the theoretical models of
stellar populations by Girardi \& Salaris (2001), and Salaris \& Girardi (2002).

RC stars have been used to derive the distance, among others, to the
Galactic center (Paczynsky \& Stanek 1998), Fornax (Rizzi et al.  2007), M33
(Kim et al.  2002), and LMC (Salaris et al.  2003), and to isolate components
of the Galaxy, such as the central bar of the Galaxy and the Canis Major accreted
dwarf galaxy, or determine gradients along the Sagittarius stream (Stanek et al. 
1997; Bellazzini et al.  2006a,b; Rattenbury et al.  2007).

   \begin{figure}
   \centering
   \includegraphics[width=8.8cm]{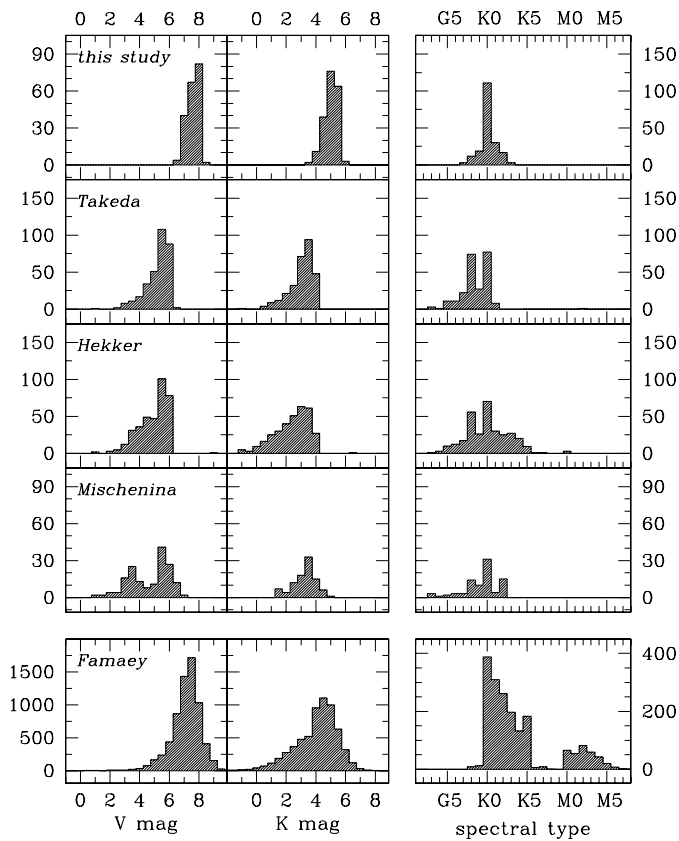}
       \caption{Distribution of $V$, $K$ and spectral type of the red clump
                stars in our survey and the surveys of Takeda et al.  (2008),
                Hekker \& Melendez (2007), Mishenina et al.  (2006), and 
                Famaey et al.  (2005).}
       \label{fig1}
   \end{figure}

A number of investigations have focused on the atmospheric properties and/or
radial velocities of bright and nearby RC stars.  McWilliam (1990) compiled
an extensive catalog of 671 G-K giants, mostly RC giants, for which he
derived $T_{\rm eff}$ from broad-band photometry, $\log g$ via isochrone
fitting (which was carried out by invoking heterogeneous calibrations of the
absolute magnitude), and individual chemical abundances by means of an LTE analysis of
high resolution spectra, while no radial velocity was measured.  Mishenina
et al.  (2006) derived both the radial velocity and the atmospheric
parameters for 177 RC stars. Values of $T_{\rm eff}$, [M/H] and elemental abundances
were derived from the classic line-by-line spectroscopic analysis (imposing
excitation and ionization equilibrium on selected FeI and FeII lines), while
$\log g$ was determined by combining the ionization balance of iron and 
fitting the wings of the CaI 6162.17 \AA\ line. Atmospheric parameters
without radial velocities were obtained by Hekker and Melendez (2007) and
Takeda et al.  (2008). They studied 366 G- and K-type giants and 322
intermediate-mass late-G giants, respectively, with many RC stars among
them, by adopting the classic line-by-line method based on FeI and FeII lines
recorded in high resolution spectra. Finally, Famaey et al.  (2005) carried
out a survey of giant stars with CORAVEL, deriving precise radial velocities
for all the K stars with $M_{\rm Hip}$$<$$2$ and M stars with $M_{\rm
Hip}$$<$$4$~mag present in the Hipparcos catalog.  Their survey provides
radial velocities, but not the atmospheric parameters, for a sample of
$\sim$ 6600 K giants, mostly RC stars.

\section{Motivation for this survey}

The afore mentioned, high-resolution spectroscopic surveys of RC stars
generally focused only on atmospheric parameters or radial velocities, but
not on both simultaneously, with the exception of the 177 RC stars observed
by Mishenina et al.  (2006).  They were also limited to the brighter RC
stars in the solar neighborhood, as illustrated in Fig.~1 by the
distribution in $V$ and $K$ magnitude of their target stars.  In addition,
they typically observed the target stars only once, so that binaries (identified on the basis of
varying radial velocities) could not be recognized and internal errors could
not be consistently evaluated by comparing of the results of different
reobservations.  To improve upon some of these shortcomings, we
conceived a new, extensive and multi-epoch high resolution spectroscopic
survey of RC stars, to which we refer in this paper as the {\em Asiago
Red Clump spectroscopic Survey} (ARCS).  This paper introduces the survey
and its data output, while a companion paper (Valentini et al.  2010, in
prep.) explores some first science results of the survey, and yet another
paper in the series (Saguner and Munari 2010, in prep.) will extend the
survey to even fainter magnitudes.

   \begin{figure}
   \centering
   \includegraphics[width=8.8cm]{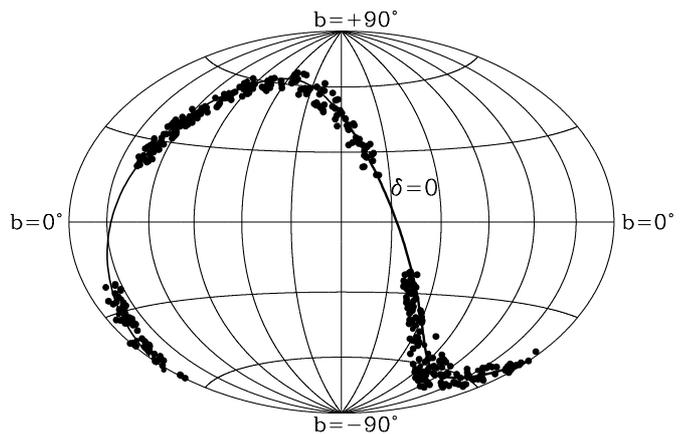}
       \caption{Aitoff projection in Galactic coordinates of our program
       stars. The thick line is the celestial equator.}
       \label{fig2}
   \end{figure}

\begin{table}
\caption{Wavelength range of the Echelle orders used in our analysis.}
\begin{center}
\begin{tabular}{ccc}
\hline   \hline 
&&\\
order & $\lambda_{\rm start}$ (\AA)& $\lambda_{\rm end}$ (\AA)\\
&&\\
38 &5831 &5965 \\
39 &5681 &5807 \\
40 &5539 &5661 \\
41 &5404 &5522 \\
42 &5275 &5389 \\
43 &5152 &5264 \\
44 &5035 &5144 \\
45 &4923 &5030 \\
46 &4816 &4922 \\ 
&&\\
\hline
\end{tabular}
\end{center}
\end{table} 

The ARCS survey observes a sample of carefully selected genuine RC
stars, distributed along the great circle of the celestial equator, away
from the Galactic plane and the effects of its extinction, fainter by $\geq$2 magnitudes
(and consequently more distant) than the surveys of Mishenina et al. 
(2006), Hekker and Melendez (2007), and Takeda et al.  (2008), and $\sim$1
mag fainter than Famaey et al.  (2005).  The magnitude distribution of the
ARCS targets is presented in Fig.~1.  An original feature of the ARCS
survey, is the derivation of the atmospheric parameters by $\chi^2$ fitting
an extensive library of synthetic spectra (that of Munari et al.  2005,
the same adopted by the RAVE survey).  The $\chi^2$ approach is the method
of choice for massive spectroscopic surveys such as the ongoing RAVE project 
with the 6dF fiber positioner at the UK Schmidt telescope in Australia, the soon-to-be
launched Gaia, the ESA's successor to Hipparcos, and the all-sky survey to  
be carried out with LAMOST (Xinglong, China). The first two surveys focus
on a limited wavelength range centered on the far-red CaII triplet, RAVE
observing at a resolving power 7500 and Gaia planned for 11500.  ARCS,
observing at twice that resolving power (20000) and over a much wider and
bluer wavelength range (4800-5900 \AA), provides an useful term of
comparison to evaluate the performances and elaborate on possible improvements. 
In addition, ARCS reobservation at a second epoch (spaced in time by at least
one month) of 1/3 of the program stars, provided that an accurate
estimate has been made of the contamination to radial velocities by binarity or
pulsations, and of the repeatibility of derived atmospheric
parameters.

\section{Target selection criteria}

The targets selected for ARCS had to meet stringent criteria designed to
provide a clean sample of genuine, unreddened RC stars, easily observable
from the northern latitudes of the Asiago observatory, and supported by
accurate spectroscopic classification and Hipparcos/Tycho-2 proper motions.

The first enforced selection criteria concerned the spectral type of the
target stars and their position on the sky.  Essentially all RC stars
present in the Hipparcos catalog have spectral types between G8III and
K2III.  We thus looked at the most complete and homogeneous set of
spectroscopic classifications, the Michigan Project.  It covers the HD stars
of the whole southern sky up to +6$^\circ$ northern declination (Houk and
Swift 2000).  From the Michigan Project catalog, we selected giants from
spectral types G8III to K2III satisfying the conditions: {\it (i)} high accuracy
of the spectral classification (quality index $\leq$2 and blank duplicity
index in the Michigan catalog), {\it (ii)} a declination within 6$^\circ$ of the
celestial equator (to facilitate observations from Asiago and to confine the
targets to a great circle on the sky), and {\it (iii)} a Galactic latitude
$|$$b$$|$$\geq$25$^\circ$ (to avoid reddening close to the plane of the
Galaxy).

We then turned to astrometric and photometric selection criteria. The ARCS
targets must also {\it (iv)} be valid entries in {\it both} the 
Hipparcos\footnote{All selection criteria involving Hipparcos information refer
to the original catalog published by ESA in 1997, and not necessarily its
revision by van Leeuwen (2007).} and Tycho-2 catalogs, {\it (v)} possess
non-negative Hipparcos parallaxes, and {\it (vi)} avoid any other Hipparcos or
Tycho-2 star closer than 10 arcsec on the sky.  Furthermore, the target
stars must {\it (vii)} be confined within the magnitude range 6.8$\leq$$V_{\rm
Tycho-2}$$\leq$8.1 (to be fainter than targets in other spectroscopic
surveys of RC stars, and at the same time still preserving meaningful
Hipparcos parallaxes), and {\it (viii)} have an absolute magnitude (from
Hipparcos parallax) incompatible with either luminosity classes V or I.  The
targets must also {\it (ix)} have a blank photometric variability flag and a
blank duplicity index in the Hipparcos catalog (to the aim of rejecting
stars affected by binarity and/or pulsations), and {\it (x)} be present in the
2MASS survey (so that their optical-IR spectral energy distribution could be
derived in combination with $B_{\rm T}$ and $V_{\rm T}$ from Tycho-2, and
($V$$-$$I$)$_{\rm C}$ from the Hipparcos catalog). 

Finally, we looked at overlaps with other spectroscopic surveys.  We {\it
(xi)} rejected stars already present in the N\"{o}rdstrom et al.  (2004)
Geneva-Copenhagen survey of dwarfs in the solar neighborhood, or included in
the radial velocity survey of giant stars by Famaey et al.  (2005).  This
set of eleven selection criteria provided a total of 381 target stars for
ARCS, from which those actually observed were randomly selected 
according to the local sideral time at the telescope. By the end of the
observing campaign, we had observed 207 ARCS targets, 66 of which (1/3 of
the total) had been observed at two distinct epochs.

Additional RC stars, brighter than the target set, were observed for
calibration purposes, to ensure commonality with other surveys, and tests. 
A total of additional 70 RC stars, with spectral type between G8III and
K2III, were observed with an instrument set-up identical to that for ARCS
targets, and for 35 of them (half of the total) a distinct second epoch
observation was in addition collected.  These calibration RC stars are
discussed in detail later in the paper, where the tests on the accuracy of
radial velocities and atmospheric parameters are described.  Of the
calibration RC stars, 10 were selected from Takeda et al.  (2008), 10 from
Hekker and Melendez (2007), 12 from Soubiran et al.  (2005), 3 from
Mishenina et al.  (2006), and 4 are members of the Praesepe open cluster.

Finally, a sample of an additional 55 non-RC stars were observed, primarly for
calibration purposes.  They include 16 well-know field stars, 24 stars from
the RAVE survey (Zwitter et al. 2008), 2 stars from Soubiran et al.  (2005),
9 members of Coma Berenices, and 4 members of Praesepe open clusters. 

\section{Data acquisition and reduction}

The target stars were observed with the REOSC Echelle spectrograph and CCD
mounted at Cassegrain focus of the Asiago 1.82m telescope.  The recorded
spectra cover the wavelength range from 3700 to 7300 \AA\ in 30 orders.  To
maximize S/N and avoid the redder wavelength region contaminated by
telluric absorptions, the data analysis was limited to the 9 adjacent
echelle orders covering the wavelength range from 4815 to 5965 \AA, 
from H$\beta$ to NaI doublet (cf Table~1).  

Being detectable only redward of 6350 \AA, fringing was not an issue.
Throughout the whole observing campaign, the slit orientation was kept fixed
to east-west and its width on the sky to 1.9 arcsec, which corresponds to a
resolving power of 20000.  The spectra were obtained as three separate and
consecutive exposures, each lasting 4 minutes.  Before reduction, the three
individual frames were median combined into a single one to automatically
reject cosmic rays and increase the S/N.  The three exposures were
autoguided to increase the sharpness in the spatial direction of the
recorded spectrum.  Along the dispersion axis, the parameters of the
autoguider were set to allow the stellar image to move for an amount
equivalent to half of the slit width, to achieve a more uniform illumination
of the entire slit width and consequently minimize any radial velocity
offset (see next section).

The spectra were reduced and calibrated with IRAF, using standard dark, bias
and dome flats calibration exposures.  Special care was devoted to the 2D
modeling and subtraction of the scattered light. Deep exposures 
of moonlight scattered by the night sky were also obtained and later
cross-correlated to the calibrated stellar spectra.  No cross-correlation
peak was found other than that expected from the stellar spectrum.  In
particular, higher velocity stars did not display a secondary peak close to
zero velocity.  These results confirmed that the sky subtraction procedure
accurately removed, from the extracted stellar spectra, the scattered moonlight. 
Therefore, any possible residual moonlight was strong enough to neither
contaminate the measurement of radial velocities, nor the determination 
of atmospheric parameters.

Exposures of a thorium lamp for wavelength calibration were obtained both
immediately before and soon after the three exposures of the star, on which
the telescope was still tracking.  These two exposures of the thorium lamp
were combined before extraction, to compensate for spectrograph flexures. 
From the start of the first thorium exposure to the end of the last, the
whole observing cycle on a program star took about 15 minutes.  According to
the detailed investigation and 2D modeling by Munari and Lattanzi (1992) of
the flexure pattern of the REOSC echelle spectrograph mounted at the Asiago
1.82m telescope, and considering that we preferentially observed our targets
when they were crossing the meridian, the impact of spectrograph flexures on
our observations corresponds to an uncertainty smaller than 0.1 km/sec, thus
completely negligible.  This is fully confirmed by (1) the measurement by
cross-correlation of the radial velocity of the rich telluric absorption
spectrum in the red portion of all our spectra, and (2) the measurement of
all night sky lines we detected in our spectra, relative to the compilations
of their wavelengths by Meinel et al.  (1968), Osterbrock and Martel (1992),
and Osterbrock et al.  (1996).

Given both the red colors of the RC stars and the instrument response, the S/N of recorded
spectra was found to steeply increase with increasing wavelength.  Over the 9
adjacent echelle orders considered for the measurement of the radial velocity
and derivation of the atmospheric parameters, the S/N
per pixel was - for all target stars - generally higher than 50 for the
bluest order and higher than 120 for the reddest one.

\begin{table}
\centering
\caption{Comparison with tabular values of the radial velocity we derived
for seven RC stars that are also IAU standard RV stars. All values are in km
s$^{-1}$.}
\begin{tabular}{ l c r c c r c c}
\hline
\hline
          &&               &          &     & &       &        \\
Star &&  \multicolumn{2}{c}{Literature} & &\multicolumn{2}{c}{Ours} & $\Delta v_{rad}$\\ 
                                                                 \cline{3-4} \cline{6-7}
     && v$_{rad}$& $\sigma_{\rm Vrad}$& & $v_{\rm rad}$ & $\sigma_{\rm Vrad}$ &\\  
                 &          &     & &       &      &   \\
HD 003712  & K0III  & $-$3.9  & 0.1 & & $-$4.1  & 0.3 &$-$0.2   \\ 
HD 012929  & K2III  & $-$14.3 & 0.2 & & $-$13.8 & 0.4 &$-$0.5   \\      
HD 062509  & K0III  & $+$3.3  & 0.1 & & $+$3.3  & 0.3 &$+$0.0   \\ 
HD 065934  & G8III  & $+$35.0 & 0.3 & & $+$34.6 & 0.5 &$+$0.4   \\ 
HD 090861  & K2III  & $+$36.3 & 0.4 & & $+$36.7 & 0.4 &$-$0.4   \\ 
HD 212943  & K0III  & $+$54.3 & 0.3 & & $+$54.4 & 0.4 &$+$0.1   \\ 
HD 213014  & G9III  & $-$39.7 & 0.0 & & $-$40.0 & 0.5 &$-$0.3   \\ 
          & &               &          &     & &       &         \\ \hline
\end{tabular}
\end{table}

\begin{table}
\centering
\caption{Similar to Table~2 comparing our radial velocities with those
of 25 F, G, and K stars taken from the RAVE second data release (Zwitter 
et al. 2008).}
\begin{tabular}{ l r c c r c c}
\hline
\hline
                         &          &     & &       &     &   \\
&  \multicolumn{2}{c}{RAVE} & &\multicolumn{2}{c}{ours} &$\Delta v_{rad}$ \\ 
                                                     \cline{2-3} \cline{5-6}
& v$_{rad}$& $\sigma_{\rm Vrad}$& & $V_{\rm rad}$ & $\sigma_{\rm Vrad}$ & \\  
              &       &     & &      &      &   \\
T4678-00087-1  &  $-$2.6 &  2.4 &&   $+$0.8 &  0.2& $+$3.4\\
T4679-00388-1  & $+$13.1 &  1.0 &&  $+$18.4 &  0.7& $+$5.3\\
T4701-00802-1  & $-$42.6 &  0.5 &&  $-$41.2 &  0.3& $+$1.4\\
T4702-00944-1  & $+$26.4 &  0.5 &&  $+$29.1 &  0.4& $+$2.5\\
T4704-00341-1  & $-$20.9 &  1.6 &&  $-$19.6 &  0.6& $+$0.7\\
T4749-00016-1  & $-$29.8 &  0.6 &&  $-$27.9 &  0.3& $+$0.9\\
T4749-00085-1  & $+$61.8 &  0.7 &&  $+$62.8 &  0.9& $+$1.0\\
T4749-00143-1  & $+$18.2 &  1.2 &&  $+$18.3 &  0.6& $+$0.1\\
T4763-01210-1  &  $-$0.4 &  0.6 &&   $+$1.2 &  0.3& $+$1.6\\
T5178-01006-1  & $-$36.1 &  0.6 &&  $-$37.4 &  0.3& $-$1.3\\
T5186-01028-1  &  $-$7.9 &  1.0 &&   $-$7.7 &  0.5& $+$0.2\\
T5198-00021-1  & $-$32.9 &  0.6 &&  $-$33.3 &  0.8& $-$0.5\\
T5198-00784-1  & $-$54.7 &  1.6 &&  $-$54.0 &  0.7& $+$0.7\\
T5199-00143-1  & $-$26.9 &  2.4 &&  $-$26.7 &  0.9& $+$0.2\\
T5201-01410-1  &  $-$1.7 &  1.0 &&   $+$0.0 &  0.9& $-$1.7\\
T5207-00294-1  & $+$25.7 &  0.9 &&  $+$24.6 &  0.6& $-$1.1\\
T5225-01299-1  &  $-$8.6 &  0.8 &&   $-$9.6 &  0.3& $-$1.0\\
T5227-00846-1  &  $-$11.3&  0.7 &&  $-$10.2 &  0.8& $+$1.1\\
T5228-01074-1  &  $-$8.6 &  1.0 &&   $-$7.5 &  0.3& $+$1.1\\
T5231-00546-1  & $-$29.4 &  2.7 &&  $-$28.2 &  0.7& $+$1.2\\
T5232-00783 1  & $-$21.0 &  0.8 &&  $-$20.1 &  0.4& $+$0.9\\
T5242-00324 1  & $-$10.9 &  3.4 &&  $-$11.9 &  0.8& $-$1.0\\
T5244-00102 1  &  $-$7.5 &  0.9 &&   $-$5.8 &  0.5& $+$1.7\\
T5246-00361 1  & $+$11.8 &  0.7 &&  $+$11.5 &  0.5& $-$0.3\\
T5323-01037 1  & $+$20.5 &  0.7 &&  $+$22.4 &  0.7& $+$1.9\\
              &       &     & &      &      &   \\ \hline
\end{tabular}
\end{table}

\section{Radial velocities}

The radial velocity of the program stars was obtained by cross-correlating
(cf. Tonry \& Davis, 1979) with the synthetic spectral library of Munari
et al. (2005), in its version matching the 20000 resolving power of our
echelle spectra.

\subsection{Continuum normalization}

In both the determination of radial velocities via cross-correlation and
atmospheric parameters via $\chi^2$ fitting, the echelle spectra had to be
accurately continuum normalized.  The procedure of continuum normalization
that we adopted had three steps.

All spectra were first manually normalized, one by one, order by order. 
The normalization blaze functions of each order proved to be fairly
constant, and by averaging them together, we obtained a mean normalization
function for each individual order.  The second step consisted of applying
the mean normalizations to the orders of all spectra.  The resulting
normalized spectra were individually inspected and compared with each other to 
identify low spatial frequency deviations of the continua from linear slopes, and -
if required - a second normalization pass was manually carried out.  The
third and final step was to automatically renormalize, with exactly the same
parameters, both the synthetic spectra and the observed spectra, as
preliminarily normalized by the two steps just described.  We selected the
synthetic spectra in their continuum normalized form, cutting out
the wavelength intervals matching those of the nine echelle orders
selected for science operations.  The adopted function for automatic
renormalization was a Legendre polynomial of 6th order with high reject of 
1.5 and low reject of 0.5.

The adopted procedure accurately placed the observed spectra onto exactly
the same normalization system as the synthetic spectra, improving the accuracy
of the $\chi^2$ fitting, for which the most important factor was the 
S/N of the observed spectra.

\subsection{Velocity measurement}

The radial velocity was derived separately for each of the 9 echelle orders
selected for science applications.  The 9 different measurements almost
always agreed to within 0.6 km s$^{-1}$, so that the internal error in the
mean was smaller than 0.2 km s$^{-1}$.  The very few cases in which the
error exceeds the 0.6 km s$^{-1}$ spread, were found to be related to
specific problems with a single echelle order, which was therefore ignored
when deriving the final velocity for that spectrum and in the subsequent
atmospheric analysis.

The zeropoint of the wavelength (and therefore velocity) scale was checked
for each spectrum by cross-correlating the telluric O$_2$ and H$_2$O
absorption spectra recorded in various echelle orders external to the 9
selected for science applications.  In particular, the $B$ (centered on
$\sim$6900) and $A$ ($\sim$ 7620 \AA) absorption systems by O$_2$, and $a$
($\sim$7200) by H$_2$O were used.  The illumination of the slit aperture by
telluric absorptions is obviously identical to that of the stellar seeing
disk, and thus traces all wavelength conversion effects more accurately than
the use of night-sky emission lines that instead illuminate
uniformly the whole slit width.  In the vast majority of cases, the
resulting radial velocity of O$_2$ and H$_2$O telluric absorption systems
was smaller than $\pm$0.25 km s$^{-1}$, and no correction was applied to the
stellar radial velocity.  In a few cases, the shift for O$_2$ and H$_2$O
lines was larger than $\pm$0.25 km s$^{-1}$, and the radial velocity of the
corresponding stellar spectrum was corrected for.  This invariably occurred
with observations characterized by particularly good seeing, when the FWHM
of the stellar image was appreciably smaller than the projected slit
width.

   \begin{figure}
   \centering
   \includegraphics[width=8.8cm]{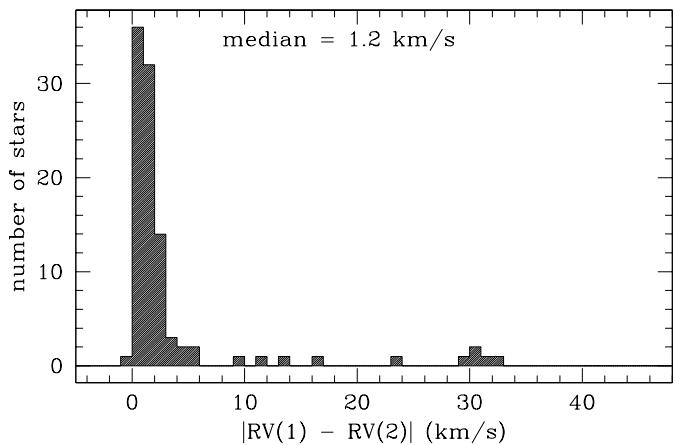}
       \caption{Distribution of the difference in velocity between two
       observations ($\geq$one month apart) for 101 program stars.}
       \label{fig3}
   \end{figure}

   \begin{figure}
   \centering
   \includegraphics[width=8.8cm]{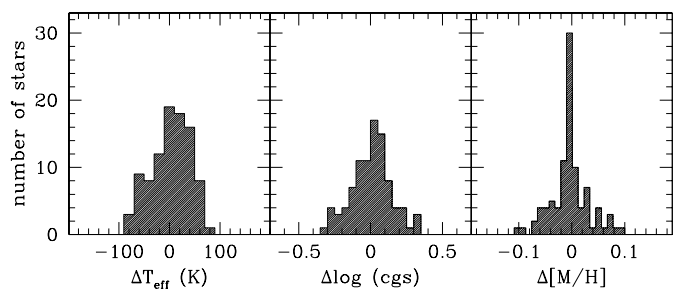}
    \caption{Differences in temperature, gravity, and metallicity between 
             reobservations of 101 program stars.}
      \label{fig4}
   \end{figure}

\subsection{Multi-epoch observations, binaries, pulsations}

As previously noted, 66 RC stars belonging to the target list and 35 of 
the calibration set were reobserved at a second epoch.  Comparison of the
radial velocities at the two distinct epochs provides both {\it (i)} an
evaluation of the internal accuracy of the radial velocities, as well as
{\it (ii)} a means of isolating binaries or stars affected by pulsations.

Figure~3 shows the histogram of the differences in radial velocity between
the two epoch observations of the 101 reobserved program stars.  The vast
majority belong to a sharp distribution characterized by a median difference
of 1.2 km s$^{-1}$, indicating that binarity and pulsations are not a
significant concern for the bulk of the RC stars selected for ARCS.  The
median value 1.2 km s$^{-1}$ is at least twice as large as the combined
internal errors, and to it some contribution must be provided by wide and
low mass ratio binaries, or low amplitude pulsations.

Ten stars are distinctive in Fig.~3 for large differences in radial velocity
of the two epoch observations.  The difference is far larger than the
observational errors, and consequently these stars are regarded as candidate
binaries or affected by large amplitude pulsations (radial modes).  We have
initiated additional observations of these ten stars to help identify the
cause of the radial velocity variability.  The results of these will be reported
elsewhere.

\subsection{Accuracy tests on IAU standards}

With the same instrument set-up and data reduction adopted for ARCS, we
observed 7 RC stars that are also IAU standard radial velocity stars
(Table~2).  The mean difference between ours and literature values is
$\Delta$RV$_\odot$= $-$0.13 km s$^{-1}$, with an rms of 0.3 km s$^{-1}$.
This confirms both the absence of a zeropoint offset in the system of ARCS
radial velocities based on cross-correlation against appropriate synthetic
template spectra, as well as their high external accuracy.

\begin{table}
\caption{Comparison between the atmospheric parameters obtained with our
$\chi^2$ method and those derived by Soubiran et al. 
(2005), Hekker and Melendez (2007), and Takeda et al. (2008) for the 32 stars
in common.}
\centering
\begin{tabular}{l@{~~~}c@{~~}c@{~~}c@{~~}c@{~~}c@{~~}c@{~~}c@{~~}c}
\hline   \hline 
         &    &     & &     &     & & &      \\
HD & T. Sp.   & \multicolumn{3}{c}{Soubiran} & &\multicolumn{3}{c}{our $\chi^2$} \\ \cline{3-5} \cline{7-9}  
         &    &   $T_{\rm eff}$& $\log g$  & [M/H]     & & $T_{\rm eff}$& $\log g$  & [M/H ]  \\ 
         &    &   (K)    & (dex)    &  (dex)      & &  (K)   & (dex)   & (dex)    \\ 
         &    &     & &     &     & & &      \\
124897   &K2III       &4208         &1.59         &$-$0.75          & & 4233     &1.60       &$-$0.78\\
161074   &K4III       &3951         &1.62         &$-$0.49          & & 3964     &1.62       &$-$0.54\\
180711   &G9III       &4751         &2.57         &$-$0.46          & & 4808     &2.55       &$-$0.46\\
212943   &K0III       &4550         &2.51         &$-$0.60          & & 4623     &2.53       &$-$0.56\\
213119   &K5III       &3845         &1.12         &$-$0.52          & & 3882     &1.15       &$-$0.58\\
216174   &K1III       &4342         &1.84         &$-$0.73          & & 4363     &1.71       &$-$0.76\\
219615   &G9III       &4795         &2.33         &$-$0.79          & & 4869     &2.27       &$-$0.75\\
005234   &K2III       &4447         &2.10         &$-$0.07          & & 4433     &2.29       &$-$0.17\\
009927   &K3III       &4343         &2.27         &$+$0.19          & & 4321     &2.15       &$-$0.19\\
010380   &K3III       &4199         &1.79         &$-$0.07          & & 4074     &1.63       &$-$0.43\\
019476   &K0III       &4852         &2.93         &$+$0.14          & & 4894     &3.18       &$-$0.14\\
039003   &K0III       &4618         &2.32         &$+$0.03          & & 4617     &2.42       &$-$0.27\\  
         &     &    &    &     & &    &    &    \\   
         &     &     & &     &     & & &      \\
         &     & \multicolumn{3}{c}{Hekker \& Melendez} & &\multicolumn{3}{c}{our $\chi^2$} \\ \cline{3-5} \cline{7-9}  
         &     & $T_{\rm eff}$& $\log g$  & [M/H]     & & $T_{\rm eff}$& $\log g$  & [M/H ]  \\ 
         &     &   (K)    & (dex)    &  (dex)      & &  (K)   & (dex)   & (dex)    \\ 
         &     &     & &     &     & & &      \\
192944   &G8III      &5000         &2.70         &$-$0.10          & &4948      &2.81       &$-$0.50   \\
203644   &K0III      &4740         &2.75         &$+$0.04          & &4708      &2.82       &$-$0.30   \\
210762   &K0III      &4185         &1.65         &$+$0.00          & &4211      &1.58       &$-$0.24   \\
214995   &K0III      &4880         &2.85         &$-$0.04          & &4703      &2.69       &$-$0.58   \\
199253   &K0III      &4625         &2.35         &$-$0.19          & &4604      &2.27       &$-$0.31   \\
213119   &K5III      &4090         &1.65         &$-$0.48          & &4061      &1.63       &$-$0.15   \\
214868   &K3III      &4445         &2.50         &$-$0.17          & &4289      &1.86       &$-$0.23   \\
215373   &K0III      &4950         &2.87         &$+$0.01          & &5019      &3.21       &$-$0.13   \\
216646   &K0III      &4600         &2.65         &$+$0.07          & &4604      &2.71       &$-$0.38   \\
219945   &K0III      &4880         &2.85         &$-$0.09          & &4800      &2.74       &$-$0.56   \\
         &      &    &    &     & &    &    &     \\   
         &    &     & &     &     & & &      \\
         &  & \multicolumn{3}{c}{Takeda} & &\multicolumn{3}{c}{our $\chi^2$} \\ \cline{3-5} \cline{7-9}  
         &      & $T_{\rm eff}$& $\log g$  & [M/H]     & & $T_{\rm eff}$& $\log g$  & [M/H ]  \\ 
         &      &   (K)    & (dex)    &  (dex)      & &  (K)   & (dex)   & (dex)    \\ 
         &    &     & &     &     & & &      \\
006186   &K0III      &4829         &2.30         &$-$0.31          & &4890      &2.48       &$-$0.50       \\
007087   &K0III      &4908         &2.39         &$-$0.04          & &4932      &2.63       &$-$0.30       \\
009057   &K0III      &4883         &2.49         &$+$0.04          & &4946      &2.87       &$-$0.24       \\
009408   &K0III      &4746         &2.21         &$-$0.34          & &4794      &2.40       &$-$0.58       \\
010761   &K0III      &4952         &2.43         &$-$0.05          & &4991      &2.64       &$-$0.31       \\
019476   &K0III      &4933         &2.82         &$+$0.14          & &5001      &2.27       &$-$0.15       \\
204771   &K0III      &4967         &2.93         &$+$0.09          & &4952      &3.24       &$-$0.23       \\
215373   &K0III      &5007         &2.69         &$+$0.10          & &5067      &3.31       &$-$0.13       \\
219945   &K0III      &4874         &2.61         &$-$0.10          & &4869      &2.75       &$-$0.38       \\
221345   &K0III      &4813         &2.63         &$-$0.24          & &4679      &2.44       &$-$0.56       \\
         &      &    &    &     & &    &    &    \\         
\hline
\end{tabular}
\end{table}

\subsection{Accuracy tests on RAVE stars}

A sample of 24 stars, well distributed in terms of luminosity class among those of 
spectral types F, G, and K, were randomly selected from the RAVE second data
release (Zwitter et al. 2008) among those most easily observable from Asiago.

We observed them with the same instrument set-up and data reduction adopted
for ARCS.  Table~3 compares the RAVE radial velocities with ours.  The mean
difference between ours and RAVE velocities is $\Delta$RV$_\odot$=$+$0.4 km
s$^{-1}$, with an rms of 1.1 km s$^{-1}$, if we ignore the first two stars
in Table~3 (one a rapidly rotating early F star, the other an M star with
emission-line cores).  Again, the comparison provides reassurance about the
quality of ARCS radial velocities, considering in particoular that the mean
accuracy of radial velocities reported in the RAVE second data release is
$\pm$1.5 km s$^{-1}$.

\section{Atmospheric parameters} 

The same continuum normalized spectra prepared for the measurement of radial
velocities were used in the derivation of the atmospheric parameters
$T_{\rm eff}$, $\log g $, [M/H], and $V_{\rm rot}\sin i $ via $\chi^2$
fitting to the synthetic spectral library of Munari et al.  (2005), which is
based on the atmospheric models by Castelli \& Kurucz (2003).  The same
synthetic library is adopted (in its version at resolving power 7500) for
the atmospheric analysis by the RAVE Survey (Steinmetz et al.  2006, Zwitter
et al.  2008).

   \begin{figure}
   \centering
   \includegraphics[width=8.8cm]{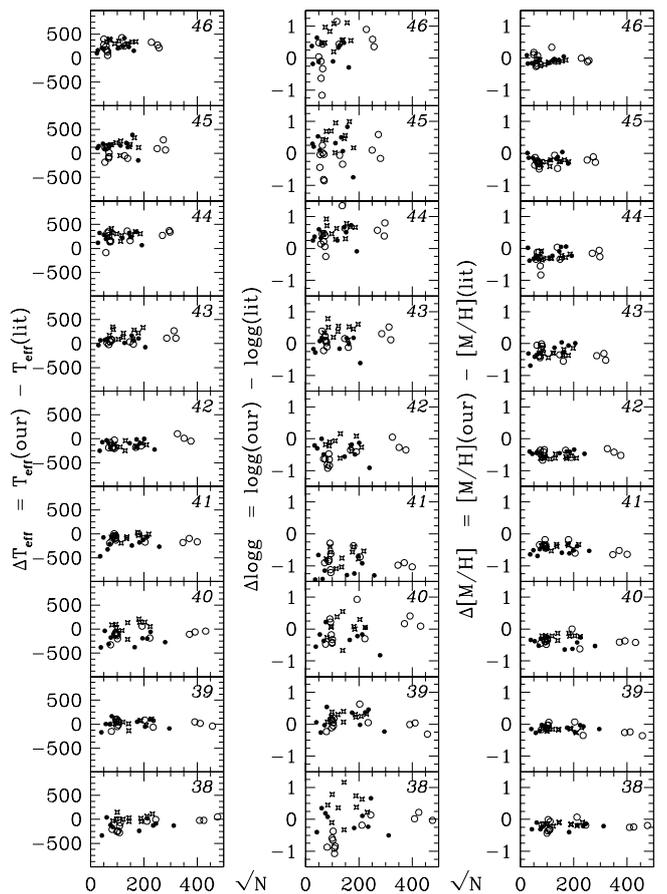}
    \caption{$\Delta$$T_{\rm eff}$, $\Delta$$\log g$, and $\Delta$[M/H]
     versus S/N for the nine Echelle orders. The plotted values are the
     differences with respect to the mean. Full circles are objects 
     from Hekker and Melendez (2007), empty circles from Soubiran
     et al.  (2005), and stars from Takeda et al.  (2008).}
      \label{fig5}
   \end{figure}

To hasten the convergence of the $\chi^2$ fitting, we considered only the
version of the synthetic library for [$\alpha$/Fe]=0.0.  Thus the
metallicity we derived refers to all metals, and is not divided into
that of the iron-peak elements separated and that of the $\alpha$ elements. 
A development we are pursuing for the ARCS survey is the derivation of individual
chemical abundances using a classical line-by-line approach, which will be
presented in a future paper, and that will naturally address the
$\alpha$-enhancement.  The Munari et al.  (2005) library adopts an
homogeneous micro-turbulence velocity of $\xi$=2 km s$^{-1}$, which is
appropriate for RC stars (as proven by the high resolution studies of 
Hekker \& Melendez 2007, Takeda et al. 2008, and Soubiran et al. 2005).

\subsection{Repeated observations}

Comparing the $\chi^2$ for the nine different Echelle orders, we obtained an
rms in the single spectrum of 47 K in $T_{\rm eff}$, 0.21 dex in $\log g$,
0.10 dex in [M/H], and 1 km s$^{-1}$ in $V_{\rm rot} \sin i$.

As discussed in Sect. 5.3, we obtained distinct observations of 101 RC stars
at two epochs separated by at least one month.  Figure~4 presents the
distributions of the differences in $T_{\rm eff}$, $\log g$, and [M/H]
between these two observations.  The distributions are quite sharp, well
centered on 0.0 and characterized by $\sigma$($T_{\rm eff}$)=39 K,
$\sigma$($\log g$)=0.24, and $\sigma$([M/H])=0.04.

\subsection{Comparison with literature data on red clump stars}

Hekker and Melendez (2007), Soubiran et al. (2005), and Takeda et al. (2008)
provide atmospheric parameters for many RC stars based on the classical
line-by-line method of imposing excitation and ionization equilibrium on
a selected sample of FeI and FeII lines.

We reobserved with exactly the same procedures and instrumental set-up
adopted for ARCS stars, a total of 10 RC stars from the list of Hekker and
Melendez (2007), another 10 RC stars from Takeda et al. (2008), and 8 RC stars
from Soubiran et al. (2005). From the latter, we reobserved an additional four
giants lying outside the G8III-K2III interval covered by ARCS targets,
bringing the total to 32 objects in common. The results are presented in
Table 4. Figure~5 shows the comparison for each echelle order individually.
It exhibits no obvious dependence on the S/N of the recorded spectrum on
the given order.

The mean differences between our results and those of Hekker \& Melendez (2007) 
for the stars in common are
               $<$$(T_{\rm our}- T_{\rm HM})$$>$=$-$45 K     ($\sigma$=76 K), 
$<$($\log g$$_{\rm our}-$$\log g$$_{\rm HM})$$>$=$-$0.05 dex ($\sigma$=0.25),
and $<$([M/H]$_{\rm our}-$[M/H]$_{\rm lit})$$>$=$-$0.26 dex ($\sigma$=0.05).
The comparison with Soubiran et al. (2005) yields
               $<$$(T_{\rm our}- T_{\rm HM})$$>$=$+$15 K     ($\sigma$=54 K), 
$<$($\log g$$_{\rm our}-$$\log g$$_{\rm HM})$$>$=$+$0.01 dex ($\sigma$=0.12),
and $<$([M/H]$_{\rm our}-$[M/H]$_{\rm lit})$$>$=$-$0.12 dex ($\sigma$=0.15).
Finally the comparison with Takeda et al. (2008) gives
               $<$$(T_{\rm our}- T_{\rm HM})$$>$=$+$21 K     ($\sigma$=62 K), 
$<$($\log g$$_{\rm our}-$$\log g$$_{\rm HM})$$>$=$+$0.15 dex ($\sigma$=0.32),
and $<$([M/H]$_{\rm our}-$[M/H]$_{\rm lit})$$>$=$-$0.27 dex ($\sigma$=0.04).

We note that there are systematic differences between the sources in
literature.  For the 147 stars in common among them, $T_{\rm eff}$ and $\log
g$ from Takeda et al.  (2008) are on average $\sim$50 K cooler and
$\sim$0.25 dex lower, respectively, than those of Hekker and Melendez
(2007), while no appreciable differences exist in terms of metallicity. 
These differences are very close to those we found, with Takeda et al. 
(2008) being 66 K cooler, 0.20 dex lighter, and the same metallicity as
Hekker and Melendez (2007).

Combining the results of the comparison with  Hekker and  
Melendez (2007), Takeda et al. (2008), and Soubiran et al. (2005), for the
32 RC stars in common, we found that
\begin{eqnarray}
<(T_{\rm our}- T_{\rm lit})>&=&-2  ~~~~~~~\pm 12    ~~~~~(\sigma=68 )~~K, \\
<(\log g_{\rm our}- \log g_{\rm lit})>&=&+0.03  ~~\pm 0.04     ~~(\sigma=0.25), \\
<([M/H]_{\rm our}- [M/H]_{\rm lit})>&=&-0.21 ~~\pm 0.03    ~~(\sigma=181). 
\end{eqnarray}

\begin{figure}
   \centering
   \includegraphics[height=8.8cm,angle=270]{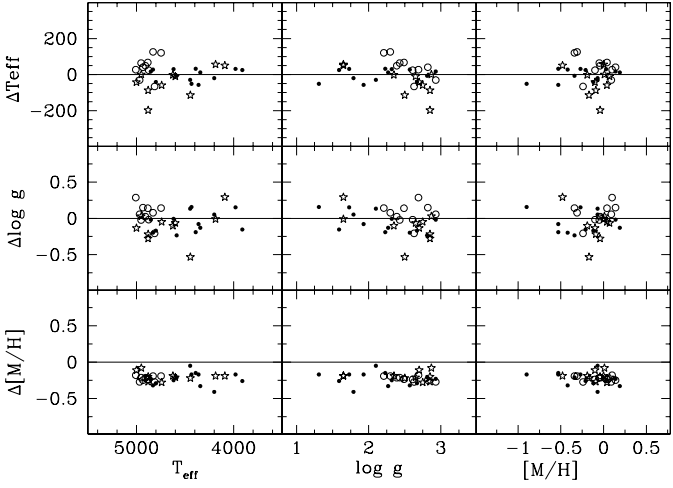}
\caption{Differences between atmospheric parameters obtained with
our $\chi^2$ technique and literature values.  Full circles are objects
from Hekker \& Melendez (1997), empty circles from Soubiran (2005),
and stars from Takeda et al.  (2008). Input data from Table~4.}
       \label{fig6}
\end{figure}

The differences in $T_{\rm eff}$ and $\log g $ are negligible, and even
smaller than the errors of the mean.  In contrast, the mean difference in
metallicity is significant.  It is similar in amount and arithmetic sign to
the mean difference between the metallicity from literature and that derived
by the RAVE survey (Zwitter et al.  2008, who also adopt a $\chi^2$ fitting
approach).

To place our results on the same atmospheric scale as that of the RC stars
in literature, {\it we therefore added +0.21 to our metallicities and
this corrected value is listed in the catalog associated with this data
release}, while no addition of an offset was required for temperatures and
gravities. 

Figure~6 plots the differences in $T_{\rm eff}$, $\log g $, and [M/H] between
our results and those of Hekker and Melendez (2007), Takeda et al.  (2008),
and Soubiran et al.  (2005) for the 32 RC stars in common.  Apart from the
rigid shift in metallicity, Fig.~6 shows that there are no systematic
trends.

\subsection{Tests on open clusters}

To check the external consistency of our atmospheric parameters
measurements, we also observed a set of stars belonging to open clusters
whose atmospheric parameters were retrieved from literature.  This
comparison confirms the results about the RC stars in the previous section.

We observed 9 members the Coma Berenices open cluster, with spectral types
from F2III to G2V.  The results are presented in Table~5 (available
electronic only).  The average difference in temperature is $<$$(T_{\rm
ARCS}-T_{\rm lit})$$>$=+12 K ($\sigma$=200 K), and in metallicity
$<$([M/H]$_{\rm ARCS}-$[M/H]$_{\rm lit})$$>$=$-$0.26 dex ($\sigma$=0.08).

We also observed four RC stars in the Praesepe open cluster. The results are
presented in Table~6 (available electronic only).  The average difference in
temperature is $<$$(T_{\rm ARCS}-T_{\rm lit})$$>$=+13 K ($\sigma$=200 K),
and in metallicity $<$([M/H]$_{\rm ARCS}-$[M/H]$_{\rm lit})$$>$=$-$0.24 dex
($\sigma$=0.15).

\onltab{5}{
\begin{table*}
\caption{Comparison between the atmospheric parameters of some members of
the Coma Berenices open cluster (Mel 111) as given in literature and derived
via $\chi^2$ fitting.
The references in Col.~6 are: (1) Wallerstein \&
Conti (1964);(2-3) Cayrel de Strobel et al.  (2001); (4) Gustafsson et al. 
(1974); (5) Claria et al.  (1996); (6) Boesgaard (1987); (7) Friel \&
Boesgaard (1992); (8) Cayrel et al.  (1988); (9) Gebran et al. (2008,).}
\centering
\begin{tabular}{l c c c l c c c c c}
\hline \hline
\multicolumn{10}{ l }{\small{}}\\
\multicolumn{10}{ l }{\small{}}\\
Star&\multicolumn{5}{ c }{Literature} &&\multicolumn{3}{ c }{our $\chi^2$}\\ \cline{2-6} \cline{8-10}
& & $T_{\rm eff}$& $\log g$& [M/H]& Ref&& $T_{\rm eff}$& $\log g$&[M/H]\\    
&   &   (K)        & (dex)  & (dex)        &      && (K) & (dex) & (dex)  \\
         &   &           &      &          &        &&  & &  \\
HD 109069& F0V  &  6864  & 4.06 &          &     9  &&7042$\pm$52&4.50$\pm$0.10&$-$0.32$\pm$0.05\\
         &   &           &      &          &        && & &   \\ 
HD 106946& F2V  &        &      &$-$0.03   &    2   &&6942$\pm$46 &4.46$\pm$0.11 &$-$0.31$\pm$0.05   \\      
         &   &      6890 &      &$-$0.031  &    6   && & &   \\
         &   &      6892 & 4.30 &          &    9   && & &   \\
         &   &           &      &          &        && & &   \\ 
HD 107611& F6V  &   6425 &      &$-$0.090  &    6   &&6543$\pm$47 &4.50$\pm$0.11 & $-$0.37$\pm$0.05  \\ 
         &   &           &      &$-$0.09   &    2   && & &   \\
         &   &      6425 &      &$-$0.056  &    7   && & &   \\
         &   &      6491 & 4.57 &          &    9   && & &   \\
         &   &           &      &          &        && & &   \\ 
HD 107793& F8V  &   6095 &      &$-$0.113  &    6   &&6159$\pm$72 &4.53$\pm$0.21 &$-$0.36$\pm$0.05   \\
         &   &           &      &$-$0.11   &    2   && & &   \\
         &   &           &      &$-$0.06   &    2   && & &   \\
         &   &      6095 &      &$-$0.059  &    7   && & &   \\
         &   &           &      &          &        && & &   \\ 
HD 107583& G0V  &        &      &$-$0.06   &    2   &&5587$\pm$59 &3.60$\pm$0.12 &$-$0.34$\pm$0.08   \\
         &   &      5960 &      &$-$0.057  &    7   && & &   \\
         &   &      5850 & 4.20 &$-$0.06   &    8   && & &   \\
         &   &           &      &          &        && & &   \\ 
HD 105863& G0V  &   5808 &      &          &    3   &&5464$\pm$47 &3.70$\pm$0.22 &$-$0.36$\pm$0.07   \\     
         &   &           &      &          &        &&              &              &                  \\ 
HD 108283& F0III  &      &      &          &        &&4939$\pm$57   &2.43$\pm$0.20 &$-$0.35$\pm$0.08   \\
         &   &           &      &          &        && & &   \\ 
HD 111812& G0III  & 4883 &      &$-$0.256  &        &&5005$\pm$42 &3.49$\pm$0.21 &$-$0.46$\pm$0.11   \\
         &   &           &      &$-$0.20   &    4   && & &   \\
         &   &           &      &          &        && & &   \\ 
HD 107700& F2III  &      &      &$-$0.10   &    2   &&6140$\pm$47 & 3.42$\pm$0.21&$-$0.39$\pm$0.11   \\
         &   &           &      &$-$0.15   &    2   && & &   \\
         &   &      6210 &      &$-$0.101  &    6   && & &   \\
         &   &      6210 &      &$-$0.148  &    7   && & &   \\ 
         &   &           &      &          &        &&  & &  \\ \hline
\end{tabular}
\end{table*}
}

\onltab{6}{
\begin{table*}
\caption{Comparison between the atmospheric parameters of some red clump
members of the Praesepe open cluster (NGC 2632) as given in literature and
derived via $\chi^2$ fitting.  The references in Col.~5 are: (1) Cayrel de Strobel et al,
(1997); (2) Taylor (1999); (3) Pasquini et al. (2000).}
\centering
\begin{tabular}{ l c c l c c c c c }
\hline \hline
\multicolumn{9}{ l }{\small{}}\\
\multicolumn{9}{ l }{\small{}}\\
Star&\multicolumn{4}{c}{Literature}& &\multicolumn{3}{c}{our $\chi^2$}\\ \cline{2-5} \cline{7-9}

 & & $T_{\rm eff}$& [M/H]& Ref&& $T_{\rm eff}$& $\log g$&[M/H]\\
 & & (K)      &(dex) &    && (K)      & (dex)   & (dex) \\
\rule[0mm]{0mm}{6mm}
HD 73665        & G8III & 4990      &$-$0.04 & 1    &&4905$\pm$52  &2.77$\pm$0.10  &$-$0.33$\pm$0.10  \\
\rule[-4mm]{0mm}{0mm}
                &       &           &$+$0.047& 2    &&  &  &  \\ 
\rule[0mm]{0mm}{6mm}
HD 73710        & G9III & 4893      &$-$0.17 & 1    &&4820$\pm$53  &2.67$\pm$0.10  &$-$0.19$\pm$0.10  \\
                &       &           &$+$0.245& 2    &&  &  &  \\ 
\rule[-4mm]{0mm}{0mm}
                &       & 4634      &      & 3      &&  &  &  \\ 
\rule[0mm]{0mm}{6mm}
HD 73598        & K0III &           &$+$0.014& 2    &&4876$\pm$53  &2.78$\pm$0.10  &$-$0.13$\pm$0.10  \\
\rule[-4mm]{0mm}{0mm}
                &       & 4799      &      & 3      &&  &  &  \\ 
\rule[-4mm]{0mm}{1cm}
HD 73974        & K0III &           &$+$0.064& 2    &&4908$\pm$61  &2.91$\pm$0.10  & $-$0.13$\pm$0.10 \\ 
\hline
\end{tabular}
\end{table*}
}

\begin{figure}
   \centering
   \includegraphics[width=8.2cm]{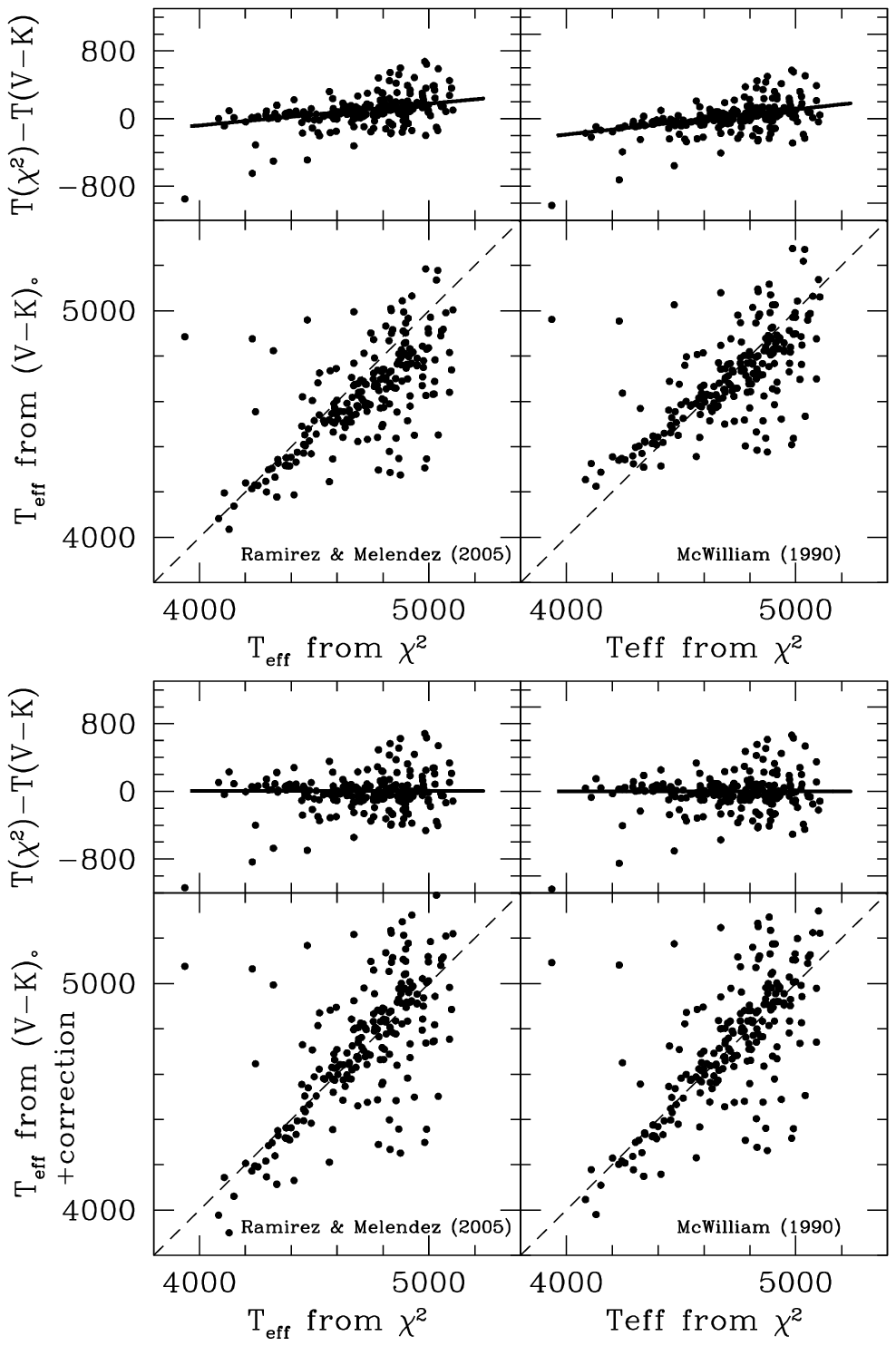}
\caption{Comparison between our temperatures and those derived from the
($V$$-$$K$) color calibrations of Ramirez \& Melendez (2005, left column)
and McWilliam (1990, right column). {\it Top panels}: using original 
$T_{\rm eff}$/($V$$-$$K$) relations. The dashed line is the 1:1 relation,
the thick solid line the least squares fit. {\it Bottom panels:} as above
but with corrected $T_{\rm eff}$/($V$$-$$K$) relations as explained in the
text (Sect 6.4 and Eqs.~4 and 5).}
       \label{fig7}
\end{figure}

\begin{figure}
   \centering
   \includegraphics[width=8.0cm]{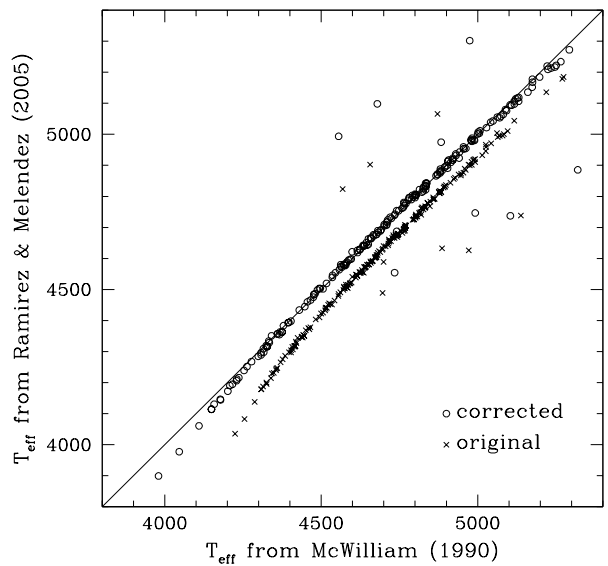}
\caption{Comparison between the Ramirez \& Melendez (2005) and
McWilliam (1990) $T_{\rm eff}$/($V$$-$$K$) relations in their
original form (crosses), and after the correction proposed in the
text (open circles).}
       \label{fig8}
\end{figure}

\subsection{Comparison with photometric temperatures}

Temperatures derived from photometric indices are usually considered in
atmospheric analysis.  Most frequently, they are used as a starting point in
the iterative analysis based on the ionization and excitation equilibrium of
FeI and FeII lines from high resolution spectroscopy (e.g.  Gratton et al. 
2006; Fulbright et al.  2007), but sometimes they are directly imported as
fixed values into abundance analysis (e.g.  McWilliam 1990; Honda et al. 
2004).  The most frequently used color index for deriving photometric
temperatures is $V$$-$$K$.  Two of the most widely adopted calibrations of
$T_{\rm eff}$ as a function of $V$$-$$K$ applicable to G-K giants (and
therefore to RC stars), were presented by Ramirez and Melendez (2005) and
McWilliam (1990).
  
Adopting the final corrected metallicity derived by our $\chi^2$ analysis,
we computed the temperature of our program stars following Ramirez and
Melendez (2005), who elaborated upon and expanded previous
extensive work by Alonso et al (1996, 1999).  The form of their relation is
complex 
\begin{displaymath}
T_{\rm eff}=5040/\theta_{\rm eff} + P\{(V-K),[Fe/H]\}, 
\end{displaymath}
where
\begin{eqnarray}
\theta_{\rm eff}&=&0.4813 + 0.2871(VK) - 0.0203(V-K)^2 \nonumber \\ 
                & &-0.0045(V-K)[Fe/H] + 0.0062[Fe/H] \nonumber \\
                & &- 0.0019[Fe/H]^2, \nonumber
\end{eqnarray}
and P\{($V$$-$$K$),[Fe/H]\} is a polynomial expression that takes different forms
over different ranges of metallicity.  Here the $V$ band is $V_{\rm T}$
Tycho-2, and $K$ is the $K_{\rm s}$ from 2MASS.  The comparison with our
temperatures is shown in the top two, left panels of Fig.~7.  The median
difference is 88~K (photometric temperatures being cooler) and there is also
a systematic trend, with photometric temperatures drifting progressively
toward cooler values as the temperature inferred from $\chi^2$ increases.

The form of the relation between $T_{\rm eff}$ and $V$$-$$K$ proposed by
McWilliams (1990) is simpler, and does not contain terms depending on
metallicity, i.e., $T_{\rm eff}$ = 8595 - 2349($V$$-$$K$) + 321($V$$-$$K$)$^2$
- 8($V$$-$$K$)$^3$, where $V$ and $K$ are standard Johnson bands.  The
comparison with our temperatures is shown in the top two right panels of
Fig.~7.  We obtained the Johnson $V$ magnitudes from $V_{\rm T}$ Tycho-2
following the transformation by Bessell (2000), while the $K$ magnitudes
were assumed to be equal to $K_{\rm s}$ from 2MASS without transformation.
The median difference from our $\chi^2$ temperatures is much smaller, this
time just 18~K (photometric temperatures being cooler), while the slope of
the least squares fitting is only 15\% steeper than in the case of
Ramirez and Melendez (2005). The comparison between $T_{\rm eff}$ derived
following Ramirez \& Melendez (2005) and McWilliam (1990) is given by
the crosses in Fig.~8, and clearly illustrates the systematic offset of 70~K
between the two calibrations.

   \begin{figure}
   \centering
   \includegraphics[width=8cm]{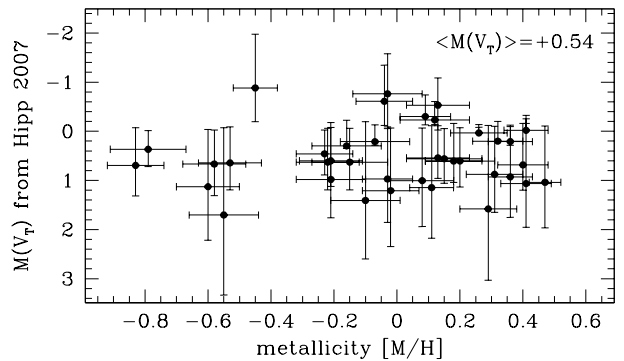}
  \caption{Absolute magnitude in the Tycho $V_{\rm T}$ band, as a function of
  metallicity, for the stars in the calibration set characterized by high
  Galactic latitude ($b$$\geq$25$^\circ$), high astrometric accuracy
  ($\sigma(\pi)$/$\pi$$\leq$15~\%, and low reddening ($E_{B-V}$$\leq$0.1).}
       \label{fig9}
   \end{figure}

For the sake of discussion, we searched for simple correction terms to the expressions of
Ramirez \& Melendez (2005) and McWilliams (1990) that would remove the
differences between them and with $\chi^2$ temperatures. The correction terms are
\begin{equation}
T_{\rm Ram Mel}^{corrected} = T_{\rm Ram Mel} - 255[(V-K) - 3.027] 
\end{equation} 
for Ramirez \& Melendez (2005), and
\begin{equation}
T_{\rm McWilliam}^{corrected} = T_{\rm McWilliam} - 305[(V-K) - 2.607]
\end{equation} 
for McWilliam (1990). The relation between these corrected temperatures and
our $\chi^2$ values is shown in the four bottom panels of Fig.~7.  The
corrections eliminate both the offsets and the trends.  This of course reflects
also the closer agreement between the Ramirez \& Melendez (2005) and
McWilliams (1990) temperatures, as illustrated by the open circles in
Fig.~8. 

That photometric temperatures are systematically cooler than spectroscopic ones
is a well documented finding. Hill et al. (2000) found a mean difference of 
$\Delta T_{\rm eff}$=150~K, Allende Prieto et al. (2004) derived 
$\Delta T_{\rm eff}$=119~K, Mishenina et al. (2006) obtained 
$\Delta T_{\rm eff}$=47~K, Hekker \& Melendez found $\Delta T_{\rm eff}$=56~K,
and Takeda et al. (2008) got $\Delta T_{\rm eff}$$\sim$50~K. We note
that Kovtyukh et al. (2006) found that photometric temperatures
derived following Alonso et al. and Ramirez \& Melendez formulations are
50~K cooler than spectroscopic ones, while the difference with McWilliam
(1990) reduces to 19~K, very similar to our results.

The corrections to Eqs. (4) and (5) have a negligible effect on the dispersion
of the points in Fig.~7, which remains high, with an rms of 195 K for both
Ramirez \& Melendez (2005) and McWilliams (1990).  This is far larger than
the rms of 61 K (and both null offset and null trend) in the difference
between our $\chi^2$ temperatures and those of Hekker and Melendez (2007),
Soubiran et al.  (2005), and Takeda et al.  (2008) for the 32 program stars
in common discussed above.  It confirms that while photometric temperatures
may be useful starting points, a far more robust determination of $T_{\rm
eff}$ is obtained by either the FeI/FeII line-by-line analysis and 
the $\chi^2$ fitting.

The larger dispersion in the points in Fig.~7 toward hotter
temperatures is caused by the Wien peak of the stellar energy
distribution moving toward shorter wavelengths.  When both bands are on the
Rayleigh-Jeans tail of the energy distribution, the $V$$-$$K$ color is less
sensitive to $T_{\rm eff}$ .  The effective wavelength of the Tycho-2
$V_{\rm T}$ band for the spectral type K0III is 5375~\AA, and that of the 
Johnson $V$ band is 5575 ~\AA\ (cf.  Fiorucci and Munari 2003).  Thus, while for
4200~K the Wien peak at 6900~\AA\ is still in-between the $V$ and $K$ bands, at
5200~K it matches the effective wavelength of the $V$ band with a consequent
rapid loss in sensitivity.

\begin{figure}
   \centering
   \includegraphics[width=8.8cm]{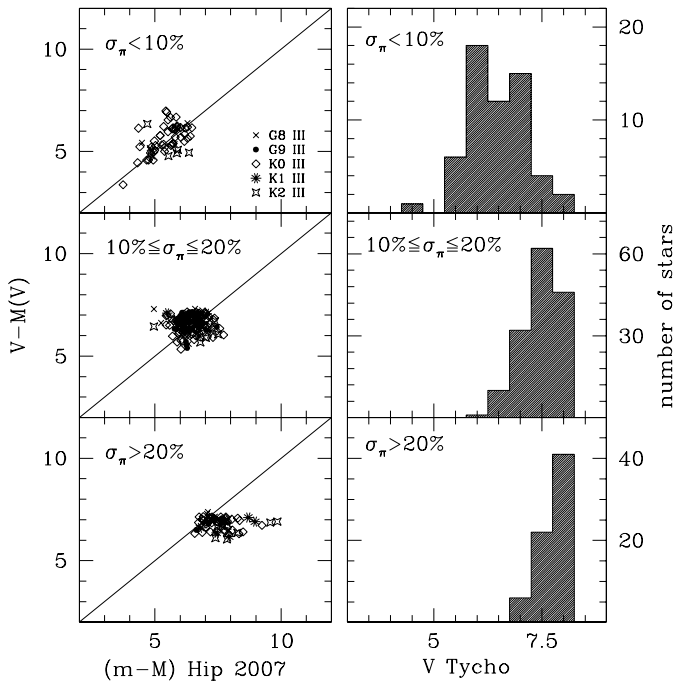}
       \caption{Left columns present, for three different bins in the error
       of parallaxes and for all our program red clump stars, a comparison
       between the distance modulus obtained with the photometric parallaxes
       from Keenan and Barnbaum (2000) calibrations and distance modulus
       obtained with the revised Hipparcos parallaxes of van Leeuwen (2008). 
       The lines represent the 1:1 correspondence.  Each different spectral
       type is represented with a different symbol.  The right columns give
       the corresponding distribution in $V_{\rm T}$ magnitudes.}
       \label{fig10}
   \end{figure}

\begin{figure}
   \centering
   \includegraphics[width=8.8cm]{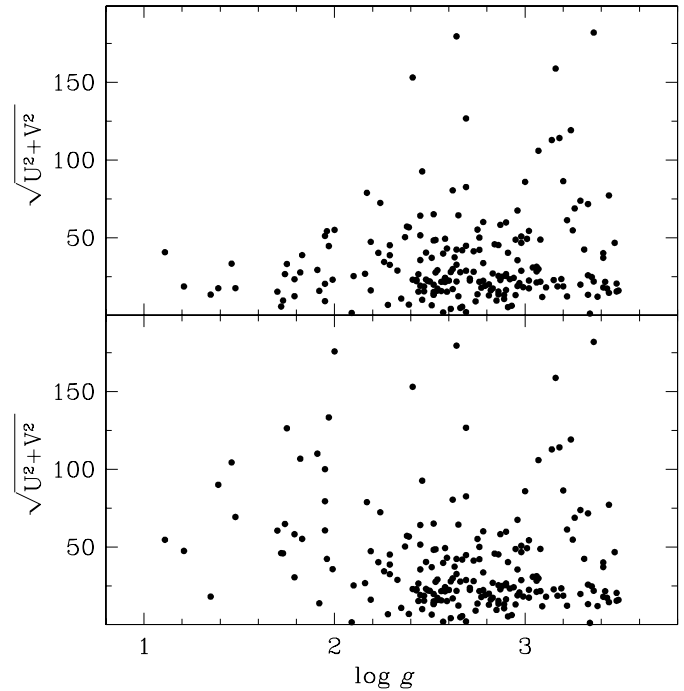}
   \caption{Relation between orbital energy and surface gravity. The top panel
         is built assuming that all stars share the same luminosity class
         III, the bottom one assuming a luminosity class II for stars with
         $\log g$$\leq$2.}
       \label{fig11}
\end{figure}

\section{Absolute magnitudes and distances}

As illustrated in Sect. 1, there is some controversy in the literature about 
whether the absolute magnitude of RC stars does or does not depend on
metallicity.  Figure~9 plots the Tycho-2 absolute magnitude M($V_{\rm T}$)
as a function of metallicity for the RC stars in our calibration set (thus
matching the same selection criteria as our target RC stars, except for
their brighter apparent magnitude) that have the most precise revised
Hipparcos parallaxes ($\sigma(\pi)$/$\pi$$\leq$15~\%) and lowest reddening
($E_{B-V}$$<$0.1).  There is no statistically significant dependence on
metallicity, and the mean absolute magnitude is $<$M($V_{\rm T}$)$>$=+0.54.

Previous works on RC stars have been based on distances derived from
Hipparcos parallaxes.  This is a valid approach as long as the error in the
astrometric parallax is small.  This is not the case for our fainter target
stars, for which the error in the Hipparcos parallax is $\sigma
(\pi)$ / $\pi$$\geq$50\%.  The corresponding uncertainity is $\geq$50\% in
the distance, and $\geq$0.88~mag in the distance modulus (m$-$M).

We therefore decided to derive uniform spectrophotometric distances for all our
program stars.  We adopted the intrinsic absolute magnitudes in the Johnson
$V$ band calibrated by Keenan \& Barnbaum (2000).  They are given separately
for the various spectral types (G8III to K2III) covered by our program
stars, and are calibrated on RC stars with the most precise Hipparcos
parallaxes.  To derive the distances, we transformed the Tycho-2 $V_{\rm T}$
into the corresponding Johnson $V$ band following Bessell (2000) relations,
and adopted the reddening derived from the all-sky mapping by Arenou et al. 
(1992).  How these distances compare with the revised Hipparcos distances
(van Leeuwen 2007) is illustrated in the left panels of Fig.~10.  There,
the various spectral types are plotted separately.  For the RC stars with
the most accurate revised Hipparcos parallaxes ($\sigma(\pi)$/$\pi$$<$10\%),
there is no offset and the distances derived adopting Keenan \& Barnbaum
(2000) absolute magnitudes are aligned along the 1:1 relation. 
This proves the validity of our choice for the reference absolute magnitudes. 
For RC stars with revised Hipparcos parallaxes of intermediate precision
(10\%$\leq$$\sigma(\pi)$/$\pi$$\leq$20\%), the correspondence is still good
even if the tail to the right in the distribution caused by the degraded
astrometric parallaxes starts to become visible.  Finally, for stars with a
revised Hipparcos parallax of low accuracy ($\sigma(\pi)$/$\pi$$>$20\%),
Fig.~10 clearly illustrates how spectrophotometric distances are more accurate
and unbiased. 

\begin{table*}
\caption{Content and description of the annexed catalog.}
\centering
\begin{tabular}{l l l l l}
\hline
\hline
  &  &  & &\\
Characters   &   Character  &Units  &Symbol & Description \\ 
             &Code           &             &             &        \\ \hline
             &               &             &             &         \\
1-8          & I8            & ...         &HD                & HD number \\
10-14        & I5            & ...         &HIP               & HIP number  \\
16-21        & I6           & ...         &TYC1           & TYCHO-2 1st identifier \\
23-27        & I5           & ...         &TYC2           & TYCHO-2 2nd identifier \\
29            & I1           & ...         &TYC3              & TYCHO-2 3rd identifier \\
31-41            & A11            & ...         &spTyp        & Spectral type from Michigan catalog \\
43-54            & F12.8           & ...         &RA         & Right ascension (J2000)\\
56-67            & F12.8           & ...         &DE         & Declination (J2000)    \\
69-76          & F8.4          & deg         &GLat          & Galactic latitude \\
78-85          & F8.4          & deg         &DLon          & Galactic longitude\\
87-92           &F6.4          & mag         & Hp           & Hipparcos $H_{\rm p}$ magnitude\\
94-99           &F6.4          & mag         & eHp          & error on $H_{\rm p}$ \\
101-104           &F4.2          & mag         & V-I          & ($V$$-$$I$)$_{\rm C}$ from Hipparcos catalog\\
106-109           &F4.2          & mag         & eV-I         & error on ($V$$-$$I$)$_{\rm C}$\\
111-117           &F7.2          & mas         & parHip       & Hipparcos (ESA 1997) parallax \\
119-124           &F6.2           &mas         & eparHip      & error on Hipparcos (ESA 1997) parallax\\
126-130            &I5             &pc           & d+          & max distance from Hipparcos (ESA 1997) parallax\\
132-136           &I5             &pc           & d-          & min distance from Hipparcos (ESA 1997) parallax\\
138-143           &F7.2          & mas         & parVL        & revised Hipparcos (van Leeuwen 2007) parallax \\
146-151           &F6.2           &mas         & eparVL       & error on revised Hipparcos (van Leeuwen 2007) parallax\\
153-158           &I6             &pc           & d+          & max distance from revised Hipparcos (van Leeuwen 2007)  parallax\\
160-165           &I6            &pc           & d-          & min distance from revised Hipparcos (van Leeuwen 2007)  parallax\\
167-172           & F6.3          &mag          &BT           & Tycho $B_{\rm T}$ magnitude \\
174-178           & F5.3          &mag          &eBT          & error on Tycho $B_{\rm T}$ magnitude    \\
180-185           & F6.3          &mag          &VT           & Tycho $V_{\rm T}$ magnitude \\
187-191           & F5.3          &mag          &eVT          & error on Tycho $V_{\rm T}$ magnitude    \\
193-198           & F6.1          &mas/yr       &pmRA         & Tycho-2 RA proper motion\\
200-203           & F4.1          &mas/yr       &epmRA        & error on Tycho-2 RA proper motion \\
205-210           & F6.1          &mas/yr       &pmDEC        & Tycho-2 DEC proper motion\\
212-215           & F4.1          &mas/yr       &epmDEC       & error on Tycho-2 DEC proper motion \\
217-222           & F6.3          &mag          & J2MASS      & 2MASS J magnitude \\
224-228           & F5.3          &mag          & eJ2MASS     & error on 2MASS J magnitude \\
230-235           & F6.3          &mag          &K2MASS       & 2MASS H magnitude \\
237-241           & F5.3          &mag          &eK2MASS      & error on 2MASS H magnitude \\
243-248           & F6.3          &mag          &K2MASS       & 2MASS K magnitude \\
250-254           & F5.3          &mag          &eK2MASS      & error on 2MASS K magnitude \\
256-258           & A3            &...          &2MASSQF      & 2MASS  quality index \\
260-265           & F6.3          &mag          &IDENIS       & DENIS $I$ magnitude \\
267-270           & F4.2          &mag          &eIDENIS      & error on $I_{\rm DENIS}$\\
272-274           & I3            &...          &DENISQF      & DENIS  quality index \\
276-291           & F16.8          & ...         & HJD         & Heliocentric Julian date of Observation \\
293-298           & F6.1          &(km s$^{-1}$)&RV           & Heliocentric radial velocity  \\
300-302           & F3.1          &(km s$^{-1}$)&eRV          & Error on heliocentric radial velocity\\
304-307           & F4.2          &(km s$^{-1}$)&TCor         & Radial velocity of telluric absorption spectrum\\
309-312           & I5            &K            &Teff         & Effective temperature \\
314-315           & I2            &K            &eTeff        & Error on effective temperature\\
317-320           & F4.2          &dex          &logg         & Surface gravity \\
322-325           & F5.2          &dex          &elogg        & Error on surface gravity \\
327-331           & F5.2          &dex          &[M/H]        & Metallicity \\
333-336           & F4.2          &dex          &e[M/H]       & Error on metallicity\\
338-341           & F4.1          &(km s$^{-1}$)& Vrot        & $V_{\rm rot} \sin i$\\
343-346           & F4.1          & (km s$^{-1}$)& eVrot      & error on  $V_{\rm rot} \sin i$\\
348-353           & F6.1          &mag           & E(B-V)     & spectro-photometric distance\\
355-359           & F5.1            &pc           & d           &  error on spectro-photometric distance \\
361-369           & F9.7            &pc           & ed          &reddening from Arenou et al. (1992)\\
371-379           & F9.4           &pc           & X          & Galactic X coordinate \\
381-389           & F9.4            &pc           & Y         & Galactic Y coordinate  \\
391-399           & F9.4            &pc           & Z         & Galactic Z coordinate  \\
401-408           & F8.3          &(km s$^{-1}$)&U            & U velocity \\
410-417           & F8.3          &(km s$^{-1}$)&V            & V velocity \\
419-427           & F8.3          &(km s$^{-1}$)&W            & W velocity \\
429-437           & I8           & ...         &BF           & flags (B= binary candidate; lcII= luminosity class II; Prae= in Praesepe cluster;\\
                  &               &             &             & Coma= in Coma cluster; TK08= in Takeda et al. (2008); IAU= standard RV star;\\
                  &               &             &             & HM07= in Hekker \& Melendez (2007); SB05= in Soubiran et al. (2005)\\
\hline
\end{tabular}
\end{table*}

The energy of the Galactic orbit of a star should obviously be unrelated to
its surface gravity (which is instead correlated with the absolute
magnitude).  To check this, we plot in the upper panel of Fig.~11 all our
program RC stars in a $\log g$ / orbital energy diagram (for derivation
of $U$,$V$,$W$ component velocities see next section).  Stars at
1$\leq$$\log g$$\leq$2 have an orbital energy systematically lower
than stars at $\log g$$\geq$2. We note that the surface gravity
of genuine RC stars is confined to $\log g$$\geq$2, once our observational
errors are properly take into account, while the surface gravity of G-K
giants of luminosity class II is 1$\leq$$\log g$$\leq$2 (cf Strai\v{z}ys
1992).  If the absolute magnitude of the 1$\leq$$\log g$$\leq$2 stars in
Fig.~11 is revised to that given by Sowell et al.  (2007) for luminosity
class II, the diagram in the lower panel of Fig.~11 is obtained.  There the
orbital energy of 1$\leq$$\log g$$\leq$2 stars is equal to that of $\log
g$$\geq$2 stars, as it should be.  Therefore, in the catalog associated with
this paper, the distance (and related quantities) for these 26 1$\leq$$\log
g$$\leq$2 stars is listed as being derived assuming for them a luminosity class II
and thus considering erroneous the luminosity class III listed in the
Michigan Project Catalog V (Houk \& Swift 2000).  These 26 stars of
luminosity class II are obviously not included in Fig.~10, which is meant to
validate the absolute magnitudes of Keenan \& Barnbaum (2000) for genuine RC
stars (thus luminosity class III).

\section{The catalog}

The results of our survey is presented in the annexed catalog, available
only in electronic form. Its format and content are illustrated in Table~7.
In addition to the survey products, the catalog also includes photometric and
astrometric information retrieved from the literature. Results from separate
observational epochs are both listed.  A conversion of the space velocity to
the usual ($U$, $V$, $W$) system by using as input data position ($\alpha$,
$\delta$), proper motion ($\mu_{\alpha}$, $\mu_{\delta}$), radial velocity,
and distance was carried out following the formalism of Johnson and Soderblom
(1987). We adopted a left-handed system with U becoming positive outwards
from the Galactic center.

The catalog is divided into three main parts, reflecting the target
selection criteria discussed in Sect.~3. It contains 433 entries, for 
332 stars, 101 of them reobserved at a second epoch.

The first part of the catalog focuses on the target stars (those plotted in
the brightness distribution histograms of Fig.~1), which are not present in
other surveys of RC and giant stars.  They are 207, and for 66 of them the
results of a second observation are given.

The second part focuses on brighter RC stars, observed for both calibration
purposes and to ensure commonality with other surveys of RC stars. There are 70
RC stars, and for half of them (35 stars), the catalog gives the results of a second epoch
observation. In this second part of the catalog, we have first listed calibration
RC stars (31 in all) not present in other survey and sharing (except for a
brighter apparent magnitude) the same selection criteria as those for the
objects included in the first part of the catalog (the target set). We have
then listed the other calibration RC stars in common with other surveys (39 stars).

Finally, the third part of the catalog lists the results for all other
55 test and calibration stars that are not RC stars.

\begin{acknowledgements}
We would like to thank R. Barbon for assistance during the realization of
the project, K.  Freeman for encouragement and useful advice, F.  van
Leeuwen for comments of revised Hipparcos distances, P.  Valisa for
assistance with the collection of some spectra, T. Saguner for useful
discussions, and especially M. Fiorucci for coding part of the $\chi^2$
pipeline we adopted.  This work has been in part supported by grant PRIN INAF
2008 (P.I.  Michele Bellazzini, INAF-OABO).
\end{acknowledgements}

\end{document}